\documentstyle [aps,epsfig,float,twocolumn,prl]{revtex}
\begin{document}
\twocolumn[\hsize\textwidth\columnwidth\hsize\csname
@twocolumnfalse\endcsname
\title{Quantum Dynamics of Spins Coupled by Electrons in 1D Channel}
\author{{\bf Dmitry Mozyrsky},$^1$ {\bf Alexander Dementsov}$^2$ and {\bf Vladimir Privman}$^2$\\
$^1$Theoretical Division, Los Alamos National Laboratory, Los Alamos, NM 87545\\
$^2$Department of Physics, Clarkson University, Potsdam, NY 13699}
\maketitle
\begin{abstract}
We develop a unified theoretical description of the induced
interaction and quantum noise in a system of two spins (qubits)
coupled via a quasi-one-dimensional electron gas in the Luttinger
liquid regime. Our results allow evaluation of the degree of
coherence in quantum dynamics driven by the induced indirect exchange
interaction of localized magnetic moments due to conduction
electrons, in channel geometries recently experimentally studied
for qubit control and measurement.
\hfill\break{\tt PACS:\ 85.85.+j, 05.30.-d, 05.60.Gg}
\end{abstract}\vphantom{\ }]

Recently, there has been much interest in coherent quantum
dynamics of coupled two-level systems (qubits) for quantum
information processing. Realizations are sought such that
qubit-qubit interactions can be externally controlled over short
time scales of quantum ``gate functions,'' in the parameter regime
ensuring that relaxation and decoherence are negligibly small over
a large number of gate cycles. There have been several proposals
for qubit systems in semiconductor heterostructures, with direct
coupling \cite{R1}, typically via shared electron wave
functions, or indirect coupling, specifically via excitations of
the conduction electron
gas \cite{R2}. In the latter approaches, the medium
that induces the indirect interaction, can also act as a ``heat
bath'' resulting in relaxation and decoherence. Usually, strongly
correlated, low-temperature conditions have been assumed \cite{R3}
in order to ensure high degree of coherence. In this work we, for
the first time, develop a unified theoretical derivation of the
Ruderman-Kittel-Kasuya-Yosida (RKKY) type \cite{R4} induced interaction
incorporating the description of relaxation effects resulting
from the electron gas ``bath.''

A recent experiment \cite{R5} on coupled quantum dots has
demonstrated the realizability of indirect interaction for control
of two-qubit dynamics. Several experimental setups \cite{R6}
suggest that a quasi one-dimensional (1D) channel geometry for the
conduction electron gas is promising for quantum measurement
required for quantum computing. Furthermore, there is experimental
evidence \cite{R7} of Luttinger liquid behavior in electron
transport in quasi 1D structures. Therefore, we are going to
consider the 1D-channel qubit-qubit coupling via indirect RKKY
interaction mediated by Luttinger liquid of electrons, and we
assume spin-$\scriptstyle{1/2}$ qubits.

It has been commonly accepted \cite{R8} that the ground state
excitations of a 1D interacting electron gas within the Luttinger
liquid model can be described by the following Hamiltonian,
\begin{equation}
H_0 =\sum_{i=c,s} {v_i\over 4\pi}\!\int\! dx \left[g_i(\partial_x\phi_i)^2 +
g_i^{-1} (\partial_x\theta_i)^2\right]\, .\label{1}
\end{equation}
The phase fields, $\phi_{c(s)}(x)$ and $\theta_{c(s)}(x)$, with
subscript indices describing charge and spin degrees of freedom,
respectively, obey commutation relations $[\partial_x
\phi_{c(s)}(x),\theta_{c(s)}(x^\prime)] = 2\pi i
\delta(x-x^\prime)$. Consequently, $\theta$ and $\partial_x \phi$
can be viewed as canonical variables. Here and in what follows we
set $\hbar=1$ and $k_{\rm B}=1$. The Hamiltonian has a simple additive structure as
a result of spin-charge separation in 1D systems, with the charge
and spin density waves of the liquid having, generally speaking,
different velocities, $v_c = v_F/g_c$ and $v_s=v_F/g_s$,
respectively, where $v_F$ is the Fermi velocity. The constant
$g_c > 0$ accounts for the electron-electron interaction and is
related to the parameters of the Hubbard model \cite{R9} as
follows: $g_c \simeq (1+U/2E_F)^{-1/2}$, where $E_F=v_F k_F/2$ is
the Fermi energy, $k_F$ is the Fermi momentum, and $U$ is the
effective interaction between the electrons, $U \sim e^2/a$, where
$a$ is the short distance cutoff, $a \sim k_F^{-1}$. Also,
we assume rotational symmetry, SU(2), in the spin space
\cite{R9,R10,R11}, which implies that $g_s=1$.

The localized magnetic moments (spins) are coupled to conduction electrons
via the contact interaction,
\begin{equation}
H_{\rm int} = \sum_j {J_j}\, {\bf s}(x_j)\cdot{\bf S}_j \,
.\label{2}
\end{equation}
Here $j=1,2$\ labels impurity spins ${\bf S}_j$ positioned at
$x_j$, $J_j$ are the exchange coupling constants, and ${\bf s}(x)$ is
the local electron spin density. The spin density can be explicitly
expressed in terms of the Luttinger phase fields, see
\cite{R12,R13},
\begin{eqnarray}
&& s_z = \frac{\partial_x\theta_s}{2\pi} +
\frac{\sigma_z}{\pi a}\cos(2k_Fx + \theta_c)\cos\theta_s,\\
\label{3} && s_{\pm} = \frac{e^{\pm i\phi_s}}{\pi
a}\left[\pm i\sigma_y\cos\theta_s + \sigma_x\cos(2k_Fx +
\theta_c)\right], \label{3.1}
\end{eqnarray}
\noindent where $\sigma_{x,y,z}$ are the Pauli matrixes. The Luttinger liquid
description of the problem is generally valid in the
``hydrodynamic limit'' of spin separations $x=|x_1 - x_2| \gg a$.

Our goal is to obtain an effective description of the dynamics of
the system of two spins, with electronic degrees of freedom
integrated out. We first consider the equilibrium partition
function of the system, defined as $Z = {\rm Tr}[\exp{(-\beta
H)}]$, where $\beta = 1/T$ is the inverse temperature and $H=H_0 +
H_{\rm int}$. The partition function, $Z$, can be expressed in
terms of the spin- and temperature-dependent effective action
${\cal S}_{\rm eff}$, see \cite{R14}, $Z = \int D{\bf S}_1
D{\bf S}_2 \exp{[-{\cal S}_{\rm eff}({\bf S}_1,{\bf S}_2)]}$. As
usual, the evaluation of the effective action can only be carried
out perturbatively in the couplings $J_j$ of the spins to the
electrons. The leading non-vanishing contribution is generated by
$(1/2)\int_0^\beta d\tau_1 d\tau_2 \langle {\cal T}\ H_{\rm
int}(\tau_1)H_{\rm int}(\tau_2)\rangle_{H_0}$, where ${\cal T}$
stands for Matsubara time ordering \cite{R15}, and the
equilibrium averaging is taken with respect to the non-interacting
Hamiltonian $H_0$. The perturbative approach is, generally
speaking, invalid at sufficiently large Matsubara timescales,
i.e., low temperatures, as evident from the perturbative RG
analysis carried out in \cite{R11,R12}. At sufficiently low
temperatures, spin dynamics results in a nontrivial strong
coupling fixed point, i.e., Kondo effect. For quantum information
processing, we are interested in qubits that retain their
localized spin-$\scriptstyle{1/2}$ indentity. Therefore, we limit
our consideration to temperatures $T>T_{\rm Kondo}$.

The resulting Matsubara action is ${\cal S}_{\rm eff} = {\cal
S}_{\rm Berry} + {\cal S}_{\rm self}$. Here ${\cal S}_{\rm
Berry}$ is the Berry action term for the spins. It will be addressed
later. Presently, we focus on the self-action for the spins, given by
\begin{eqnarray}
&& {\cal S}_{\rm self} = {J^2\over 2\beta}\sum_{i\omega_n} \Big[
\chi(\omega_n, 0){\bf S}_1(i\omega_n)\cdot{\bf
S}_1(-i\omega_n)+\nonumber \\&& \chi(\omega_n, x){\bf
S}_1(i\omega_n)\cdot{\bf S}_2(-i\omega_n) + ({\bf S}_1
\leftrightarrow {\bf S}_2)\Big]\, .\label{4}
\end{eqnarray}\noindent
In (\ref{4}), $\chi(\omega_n, x)$ is the spin-spin
correlation function of the electron gas (in the imaginary time
represenation), and $\omega_n = 2\pi{}n/\beta$ are the Matsubara
frequencies. For simplicity, in (\ref{4}) and in the following
we assume that $J_1 = J_2 = J$. The correlation function
$\chi(\omega_n, x)$, which generally speaking is a tensor,
$\langle s_i(-i\omega_n,x)s_j(i\omega_n,0)\rangle$, in the SU(2)
symmetric case reduces to a scalar function, $\chi(\omega_n, x)
= \langle s_z(-i\omega_n,x)s_z(i\omega_n,0)\rangle$. It can be
evaluated by using (\ref{1}) and (\ref{3}). Assuming that the
electron gas is dense to satisfy $E_F \gg \beta^{-1}$, one obtains
(with $g \equiv g_c$),
\begin{eqnarray}
&& \chi (\omega_n, x) = -{|\omega_n|\over 4\pi
v_F^2}\exp{(-{|\omega_n x|\over v_F})}+ \nonumber \\ \nonumber
\\&& {\cos{(2k_Fx)}\over (2\pi)^{g+1}a^{1-g}}\int {d\tau\exp{(-i\omega_n\tau)}\over
(x^2+v_F^2\tau^2)^{1/2}(x^2+v_F^2\tau^2/g^2)^{g/2}}  \, .\label{5}
\end{eqnarray}\noindent
The correlation function in (\ref{5}) contains two
contributions, one due to forward scattering, peaked at wavevector
$q=0$, and another due to backwards scattering, peaked at $q=\pm
2k_F$. As expected, the forward scattering contribution is $g$-independent.

In the perturbative regime considered, the dynamics of the
spins is slow and controlled by small parameter $J^2/v_F^2$.
Therefore, we can use the small-frequency asymptotic form of the
correlation function $\chi(\omega_n, x)$ for
$|\omega_n x|/v_F \ll 1$,
\begin{eqnarray}
&& \chi (\omega_n, x) \simeq {C_1(g)\cos{(2k_Fx)}\over v_F
a^{1-g} x^g} - \nonumber
\\&&{1\over 4\pi v_F^2} \left[|\omega_n| + {C_2(g)\cos{(2k_Fx)}|\omega_n|^g\over
v_F^{g-1}a^{1-g}}\right]\, ,\label{6}
\end{eqnarray}\noindent
where $C_1 = (2\pi)^{-g-1}g^g\int dz
(1+z^2)^{-1/2}(g^2+z^2)^{-g/2}$ and $C_2 =
4(2\pi)^{-g}g^{g-1}\Gamma(1-g)\sin{[(\pi/2)(1-g)]}$. The first
term in (\ref{6}) corresponds to interaction
between the spins. The interaction is oscillatory and decays as a
power law $x^{-g}$. This result is consistent with \cite{R13}.

The second, $\omega_n$-dependent term in (\ref{6}) corresponds to
relaxation of the spins. To demonstrate this property,
we perform a transition to the real time
dynamics of the spins according to analytical continuation
rule, see \cite{R16}. We introduce a standard Keldysh contour
with forward and return branches time-ordered and
anti-time-ordered, respectively. In the Keldysh representation, the
effective action, (\ref{4}), reads
\begin{eqnarray}
&&{J^2\over 2}\int\frac{d\omega}{2\pi}\Big[{\bf {\bar
S}}_1^T(\omega){\hat{\bf \chi}}(\omega, x){\bf {\bar
S}}_1(-\omega) + {\bf{\bar S}}_1^T(\omega){\hat {\bf
\chi}}(\omega, 0){\bf{\bar S}}_2(-\omega)\nonumber
\\&& +({\bf{\bar S}}_1\leftrightarrow {\bf{\bar S}}_2)\Big]\, .\label{7}
\end{eqnarray}
\noindent Here ${\bf {\bar S}}_{i=1,2}$ are two-element column vectors, with
spin-vector operators as elements, such that their transposes are ${\bf
\bar S}^T_i = ({\bf S}_{ic},{\bf S}_{iq})$. The
``classical,'' ${\bf S}_{ic}$, and ``quantum,'' ${\bf S}_{iq}$,
components of the spin ${\bf S}_i$ are calculated as Fourier transforms
of the following combinations of spin operators on the forward ($f$) and return ($r$)
branches of the Keldysh contour,
${\bf S}_{i\,c} = ({\bf S}_{if} + {\bf S}_{ir})/2$ and ${\bf S}_q = {\bf
S}_{if} - {\bf S}_{ir}$. The response (correlation) function
${\hat {\bf \chi}}(\omega, x)$ is then a $2\times 2$ matrix, which
can be expressed in terms of the retarded and advanced response
functions, $\chi^R$ and $\chi^A$,
\begin{eqnarray}
{\hat {\bf \chi}} =\left( \matrix{ 0 & \chi^A \cr \chi^R &
\chi^K} \right) \, .\label{8}
\end{eqnarray}\noindent
The retarded and advanced response functions are related to the
Matsubara response function via the analytic continuation
$i\omega\rightarrow\omega \pm i\delta$. In thermal equilibrium $\chi^K$ can be expressed in terms of the
retarded and advanced components by using the
fluctuation-dissipation theorem,
$\chi^K = \coth{(\beta\omega/2)}(\chi^R - \chi^A)$.

Let us consider first the noninteracting case, $g=1$,
and later we will extend the results to $g \neq 1$. For
noninteracting electrons, the response function corresponds to an
Ohmic heat bath, with $C_1(1) = 1/(4\pi)$ and $C_2(1)=1$ in
(\ref{6}). Upon Fourier transform, (\ref{7}) yields
several terms. Those containing products ${\bf S}_{i\,c}\cdot{\bf
S}_{j\,q}$ represent interaction between spins, while the ${\bf {\dot
S}}_{i\,c}\cdot{\bf S}_{j\,q}$ and ${\bf S}_{i\,q}\cdot{\bf S}_{j\,q}$ terms
are responsible for energy dissipation and pure dephasing (decoherence),
respectively. The dissipative (time-derivative) terms are
small and can be neglected here. Indeed, since
${\bf {\dot S}}_i \sim  O(J^2/v_F^2)$, the ${\bf {\dot S}}$-dependent terms in
the action in (\ref{7}) are of order $ J^4/v_F^4$. The resulting
dynamics of the spins is governed by the
action ${\cal S}^K_{\rm self} =
{\cal S}^K_{\rm int}+{\cal S}^K_{\rm dec}$, where
\begin{eqnarray}
&&{\cal S}^K_{\rm int} = {\cal J}_{\rm eff}\int dt \left[{\bf
S}_{1\,c}(t)\cdot {\bf S}_{2\,q}(t) + {\bf S}_{1\,q}(t)\cdot {\bf
S}_{2\,c}(t)\right] \,, \label{9a}
\end{eqnarray}
\begin{eqnarray}
&&{\cal S}_{\rm dec}^K = {i\gamma\over 2\pi} \int dt_1 dt_2
\int_0^\infty d\omega\, \omega\cos{[\omega(t_2-t_1)]}\times
\nonumber
\\&& \coth{\left({\beta\omega\over 2}\right)}\Big\{2{\bf S}_{1q}(t_1)\cdot
{\bf S}_{1q}(t_2) - [1+\cos(2k_Fx)]\times\nonumber
\\&&{\bf S}_{1q}(t_1)\cdot {\bf S}_{2q}(t_2) + [{\bf S}_{1q}\leftrightarrow{\bf S}_{2q}]
\Big\} \,, \label{9b}
\end{eqnarray}
with
\begin{equation}
{\cal J}_{\rm eff}=J^2{\cos{(2k_Fx)}\over 4\pi v_F x}\qquad {\rm and} \qquad \gamma =
{J^2 \over 2\pi v_F^2}\; .
\end{equation}
Equation (\ref{9a}) corresponds to coherent interaction of
the spins according to the scalar ${\bf S}_1\cdot{\bf S}_2$
coupling. Equation (\ref{9b}) represents quantum noise
resulting from thermal fluctuations of the Luttinger liquid. The
noise, generally speaking, is colored. Moreover, the noises
experienced by the two spins are correlated.

The action in (\ref{9b}) can be simplified further if we
recall that the dynamics of the spins is slow, i.e., ${\bf
S}_{iq}(t_2) = {\bf S}_{iq}(t_1) + O(J^2/v_F^2)$. Then, the
self-action can be replaced by an instantaneous action, by
setting ${\bf S}_{iq}(t_1)={\bf S}_{iq}(t_2)$ in (\ref{9b}),
which now corresponds to the white noise source. The full action
of the spin system can now be written as
\begin{equation}
{\cal S}^K = {\cal S}_{\rm Berry}^K - \int dt H^K({\bf S}_{1},{\bf
S}_{2})\, ,\label{10}
\end{equation}
\noindent where the Berry action plays the role of a
ve\-lo\-city{}$\,\times\,${}mo\-men\-tum term in the
Largangian-Hamiltonian transformation~\cite{R17}. The ``Keldysh
Hamiltonian'' in (\ref{10}) is
\begin{eqnarray}
&&H^K = - {\cal J}_{\rm eff}\left({\bf S}_{1f}\cdot{\bf S}_{2f} - {\bf
S}_{1r}\cdot{\bf S}_{2r}\right) \nonumber
\\&&-4i\gamma T \Big[(\sin{\phi}\,{\bf S}_{1q} - \cos{\phi}\,{\bf
S}_{2q})^2 + ({\bf S}_{1q}\leftrightarrow{\bf S}_{2q})\Big]\,
,\label{11}
\end{eqnarray}
\noindent where $0 \leq \phi \leq \pi/2$ is defined by $\sin{(2\phi)}=
\cos^2{(k_Fx)}$.

The density matrix of the two-spin system evolves according to the
``Schrodinger equation'' ${\dot\rho} =-i{\cal T}_K (H^K\rho)$,
where Keldysh time-ordering implies that the ``forward''
operators, with subscript $f$, are positioned to the left of the density
matrix, while the ``return'' operators, labelled by $r$, are to the right of the density matrix,
\begin{eqnarray}
&&{\dot\rho} = i{\cal J}_{\rm eff}\left[{\bf S}_1\cdot{\bf S}_2,
\rho\right] - 4\gamma T\times\nonumber
\\&&
 \sum_{\alpha=x,y,z}\Big( \big[\sin \phi S_1^\alpha
-\cos{\phi} S_2^\alpha,\big[\sin \phi S_1^\alpha -\cos{\phi}
S_2^\alpha,\rho\big]\big]\nonumber
\\&&+({\bf S}_{1}\leftrightarrow{\bf S}_{2})
\Big) \, .\label{13}
\end{eqnarray}
This expression contains both the mediated interaction
and dephasing due to the electron environment.

\begin{figure}
\centering
\includegraphics[width = 8cm]{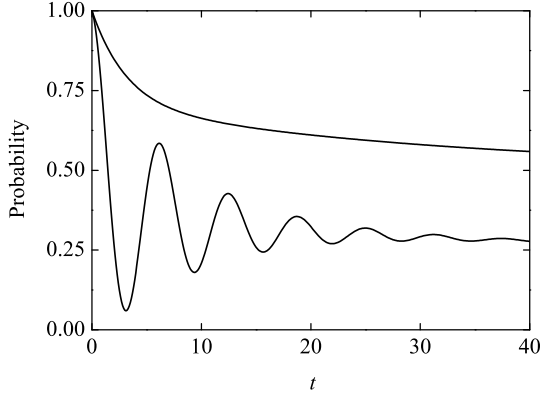}
\caption{Lower curve: The probablity, $\rho_{ |{}\uparrow \downarrow \rangle ,  |{}\uparrow \downarrow \rangle }$, to find the two-spin system in the
$|{\!\!}\uparrow \downarrow \rangle$ state as a function of time,
for a convenient set of parameter values, ${\cal J}_{\rm eff}=1$, $\gamma T
= 0.0125 $, $k_Fx=\pi $. Upper curve: The sum of the probabilities to find the two-spin system in the states $|{\!\!}\uparrow \downarrow \rangle$ or $|{\!\!}\downarrow \uparrow \rangle$, given by $\rho_{ |{}\uparrow \downarrow \rangle ,  |{}\uparrow \downarrow \rangle } + \rho_{ |{}\downarrow \uparrow \rangle ,  |{}\downarrow \uparrow \rangle } $. The large-time limiting values of these probabilities are $\scriptstyle{1/4}$ and $\scriptstyle{1/2}$, respectively.} \label{Figure1}
\end{figure}

\begin{figure}
\centering
\includegraphics[width = 8cm]{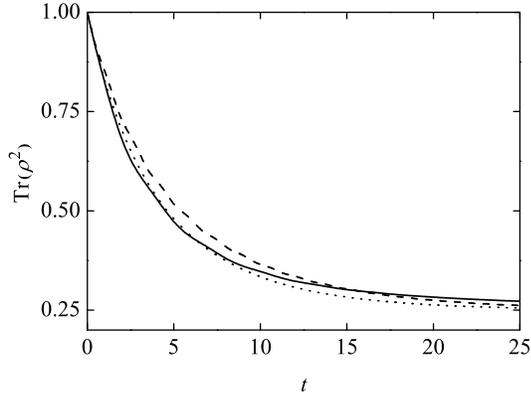}
\caption{Solid line: The quantity ${\rm Tr}(\rho^2)$, with the initial value
of 1 corresponding to a pure state, as a function of time for the
same parameter values as in Fig.~\ref{Figure1}, with positive
${\cal J}_{\rm eff}=1$. Dashed line: Parameter values illustrating the case of negative ${\cal J}_{\rm eff}=-2$, with $\gamma T
= 0.0125 $, $k_Fx=\pi /2 $. Dotted line: The deviation form a pure state (the quantum noise effect) is present even when the leading-order induced interaction is ${\cal J}_{\rm eff}=0$, with $\gamma T
= 0.0125 $, $k_Fx=\pi /4 $. The curves for ${\cal J}_{\rm eff} \neq 0$ have weak oscillations superimposed on the decay. The large-time limiting values are $\scriptstyle{1/4}$.}\label{Figure2}
\end{figure}

It is instructive to illustrate the dynamics of the two-spin
system, described by (\ref{13}), for a particular initial state of
the system, $|{\!\!}\uparrow \downarrow \rangle$. Without the
quantum noise term, proportional to $\gamma T$ in (\ref{13}), the
${\bf S}_1\cdot{\bf S}_2$ interaction would split the singlet and
triplet spin states. As a result, the system would oscillate
between the $|{\!\!}\uparrow \downarrow \rangle$ and
$|{\!\!}\downarrow\uparrow  \rangle$ states with frequency
determined by the singlet-triplet energy gap. The effects of the
noise include damping of these oscillations, as illustrated in
Fig.~\ref{Figure1}. Furthermore, for $t>0$, the system subject to
the noise will no longer remain in a pure quantum state. The
departure of the resulting mixed state from a pure state can be
measured by deviation of ${\rm Tr}[\rho^2(t)]$ from the pure-state
value of 1, as illustrated in Fig.~\ref{Figure2}. We point out
that effective evolution equations that involve only commutators,
linear in the density matrix, on the RHS, typically fail to
reproduce thermal equilibrium at large times. Instead, as seen in
Figs.~\ref{Figure1} and \ref{Figure2}$\,$---$\,$note the
assymptotic values{}$\,$---$\,$the fully random mixed state is
obtained. Therefore, the present approximation should not be used
beyond the relaxation time defined by (\ref{13}), namely, it only
applies for $t < 1/[T(J^2/v_F^2)]$, and the theory is therefore
applicable in the regime of interest for quantum computing
applications, for short and intermediate times, because both
factors in the denominator, $T$ and $J^2/v_F^2$, are small.
Similar results have been obtained in Ref.~\cite{R18}, studying
mediated interaction and decoherence due to noninteracting
electron gas.

Finally, we extend our results to the interacting
case, $g \neq 1$. Modification of the interaction term, (\ref{9a}), is
straightforward and comes from the first term on the RHS of
(\ref{6}). Equation (\ref{9a}) applies with
\begin{equation}
{\cal J}_{\rm eff}(g) = C_1(g) J^2 \frac{\cos(2k_Fx)}{v_F
a^{1-g}x^g\;}.
\end{equation}
Modification of the decoherence term, (\ref{9b}), is less obvious.
Consider for simplicity a single spin situation, when the first
term on the RHS of (\ref{6}) can be ommited, and in the second term
we can put $x=0$. The $|\omega_n|$ term then
in the brackets in (\ref{6}) produces the same, $g$-independent,
contribution to the decoherence rate as in the noninteracting
case. The $|\omega_n|^g$ term in the brackets requires careful
consideration. Indeed, we are primarily interested in the
repulsive, $U>0$, Hubbard-model interaction, i.e., $0 < g < 1$.
The two-spin contribution them yields a divergence: After setting
${\bf S}_{iq}(t_1)={\bf S}_{iq}(t_2)$ in (\ref{9b}) and
integrating over $t_1-t_2$, the frequency integral $\int
d\omega\omega^g \delta(\omega)\coth{(\beta\omega/2)}$ is divergent
for $g<1$. The origin of this divergence is actually not in the
instantaneous assumption but instead it can be traced back to too
liberal a use of the large-$\beta$ approximation in our evaluation
of the response function in (\ref{5}), and specifically, extending
the limits of the $\tau$ integration from $-\infty$ to $\infty$,
while we should have integrated from $-\beta$ to $\beta$. One can
show that $\delta(\omega)$ should be regularized as, e.g.,
$\sin^2(\beta\omega)/(\pi\beta\omega^2)$, which eliminates the
unphysical small-frequency divergence. As a result one obtains the
dephasing rate
\begin{equation}
\tau_{\rm dec}^{-1} = {4\gamma} \left(T+{C_3(g)T^g\over
v_F^{g-1}a^{1-g}}\right) \, ,\label{14}
\end{equation}\noindent
where $C_3\sim 1$ for $g\sim 1$. Equation (\ref{14}) represents
modification of the Korringa law for spin relaxation in the
interacting 1D electron gas. A precise determination of the
coefficient $C_3$ is out of the scope of the present work and will be
dealt with in a subsequent analysis based on a fully
non-equilibrium description of the system.

In summary, our main result, (\ref{13}), is the first
theoretically derived dynamical equation that incorporates both
the coherent RKKY-type induced interaction and the effects of
quantum noise due to interacting electrons in 1D conduction
channel. The approximations and assumptions involved, limit our
results to temperatures in the range $T_K < T \ll T_F$.
Furthermore, to have the noise term small, the temperature should
be actually in $T_K < T < v_F/x$, where $x$ is the qubit (spin)
separation: restoring the constants earlier set to 1, the upper
bound here is $\hbar v_F/ k_{\rm B} x$. The relative strength of
the interaction vs.\ noise terms can be also controlled by
positioning of the qubits, owing to the oscillatory dependence on
$x$, typical for RKKY-coupled systems.

The authors acknowledge helpful discussions with D.~F.~James,
I.~Martin and A.~Shnirman. The work was supported by US DOE, and by
the NSF, grant DMR-0121146.

 \vfill\eject


\begin{references} {\frenchspacing
\bibitem{R1} D. Loss and D. P. DiVincenzo, Phys. Rev. A {\bf 57}, 120 (1998);
A. Imamoglu, D. D. Awschalom, G. Burkard, D. P. DiVincenzo, D. Loss, M. Sherwin and A. Small, Phys. Rev. Lett. {\bf 83}, 4204
(1999); R. Vrijen, E. Yablonovitch, K. Wang, H. W. Jiang, A. Balandin, V. Roychowdhury, T. Mor and D. P. DiVincenzo, Phys. Rev.
A {\bf 62}, 012306 (2000).
\bibitem{R2} V. Privman, I. D. Vagner and G. Kventsel, Phys. Lett. A {\bf 239}, 141 (1998); D. Mozyrsky, V. Privman and M. L. Glasser, Phys. Rev. Lett. {\bf 86}, 5112 (2001); D. Mozyrsky, V. Privman and I. D. Vagner, Phys. Rev. B {\bf 63}, 85313 (2001); C. Piermarocchi, P. Chen, L. J. Sham and D. G. Steel, Phys. Rev. Lett. {\bf 89}, 167402 (2002); G. Ramon, Y. Lyanda-Geller, T. L. Reinecke and L. J. Sham, Phys. Rev. B {\bf 71}, 121305 (R)
(2005); M. G. Vavilov and L. I. Glazman, Phys. Rev. Lett. {\bf 94}, 086805
(2005).
\bibitem{R3} V. Privman, D. Mozyrsky and I. D. Vagner, Comp. Phys. Comm.{\bf 146}, 331
(2002).
\bibitem{R4} C. Kittel, {\sl Quantum Theory of Solids\/} (Wiley, NY, 1987).
\bibitem{R5} N. J. Craig, J. M. Taylor, E. A. Lester, C. M. Marcus, M.
P. Hanson and A. C. Gossard, Science {\bf 304}, 565 (2004).
\bibitem{R6} M. Xiao, I. Martin, E. Yablonovitch and H. W. Jiang, Nature {\bf 430}, 435 (2004); J. M. Elzerman, R. Hanson, L. H. Willems van Beveren, B. Witkamp, L. M. K. Vandersypen and L. P. Kouwenhoven, Nature {\bf 430}, 431 (2004);
M. R. Sakr, E. Yablonovitch, E. T. Croke and H. W. Jiang, e-print cond-mat/0504046.
\bibitem{R7} M. Hilke, D. C. Tsui, L. N. Pfeiffer and K. W. West,
J. Phys Soc. Jpn. {\bf 72} Suppl. A, 92
(2003); A. M. Chang, L. N. Pfeiffer and K. W. West, Phys. Rev. Lett. {\bf 77}, 2538 (1996).
\bibitem{R8} F. D. M. Haldane, J. Phys. C, {\bf 14}, 2585 (1981);
J. S\'{o}lyom, Adv. Phys. {\bf 28}, 201 (1970).
\bibitem{R9} N. Nagaosa, {\sl Quantum Field Theory in Strongly Correlated Electronic Systems},
(Springer-Verlag, Berlin, 1999).
\bibitem{R10} H. J. Schulz, in {\sl Mesoscopic Quantum Physics},
edited by Akkermans {\it et al.}, Les Houches Session LXI
(Elsevier, Amsterdam, 1995).
\bibitem{R11} C. L. Kane and M. P. A. Fisher, Phys. Rev. B {\bf 46}, 15233 (1992).
\bibitem{R12} D. H. Lee and J. Toner, Phys. Rev. Lett. {\bf 69},
3378 (1992).
\bibitem{R13} R. Egger and H. Schoeller, Phys. Rev. B {\bf 54},
16337 (1996).
\bibitem{R14} N. Nagaosa, {\sl Quantum Field Theory in Condensed Matter
Physics\/} (Springer-Verlag, Berlin, 1999).
\bibitem{R15} G. D. Mahan, {\sl Many-Particle Physics\/} (Kluwer, NY, 2000).
\bibitem{R16} A. Kamenev, e-print cond-mat/0412296.
\bibitem{R17} A. Shnirman, Z. Nussinov, J.-X. Zhu, A. V. Balatsky and Y.
Makhlin, Low Temp. Phys. {\bf 30} , 629 (2004) [Fiz. Niz. Temp.
{\bf 30} , 834 (2004).]
\bibitem{R18} Y. Rikitake and H. Imamura, Phys. Rev. B {\bf 72},
033308 (2005).}\end{references}
\end{document}